\input{aipcheck}

\documentclass[
    ,final            
  ]
  {aipproc}

\layoutstyle{6x9}

\newcommand{\etal}{{\em et al.}}

\begin{document}

\title{SM Precision Constraints at the LHC/ILC}


\author{Jens Erler\thanks{Talk presented at the 
{\sl VI Latin American Symposium on High Energy Physics} (SILAFAE 2006), 
Hotel Krystal, Puerto Vallarta, Mexico, Nov.\ 1--8, 2006.}
}
{address={Departamento de F\'isica Te\'orica, Instituto de F\'isica, 
Universidad Nacional Aut\'onoma de M\'exico, M\'exico D.F. 04510, M\'exico}}

\begin{abstract}
The prospects for electroweak precision physics at the LHC and the ILC are 
reviewed. This includes projections for measurements of the effective $Z$ pole 
weak mixing angle, $\sin^2\theta_W^{\rm eff.}$, as well as top quark, $W$ 
boson, and Higgs scalar properties. The upcoming years may also see very 
precise determinations of $\sin^2\theta_W^{\rm eff.}$ from lower energies.
\end{abstract}

\maketitle


\section{Introduction}
\begin{figure}
  \includegraphics[height=.34\textheight]{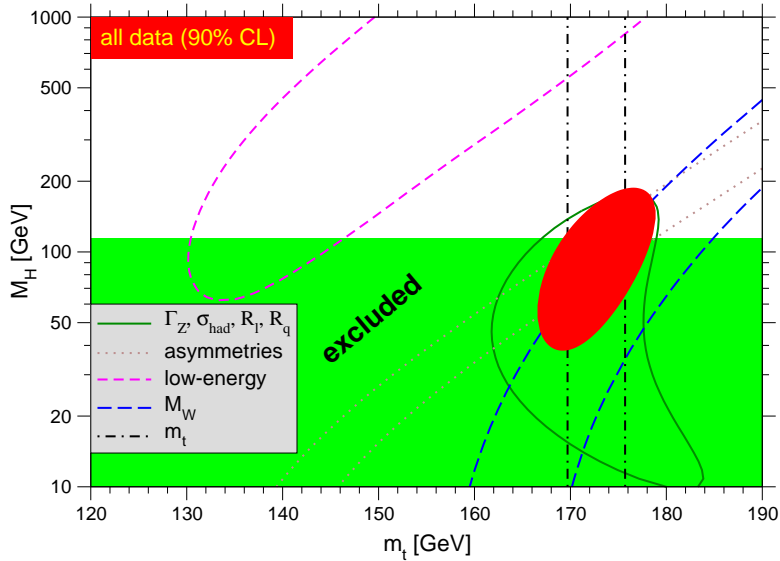}
  \caption{One-standard-deviation (39.35\%) uncertainties in $M_H$ as 
           a function of $m_t$ for various inputs, and the 90\% C.L.\ region 
           allowed by all data. The 95\% C.L.\ direct lower limit from LEP~2 is
           also shown.}
  \label{fig:mhmt}
\end{figure}

\begin{figure}
  \includegraphics[height=.34\textheight]{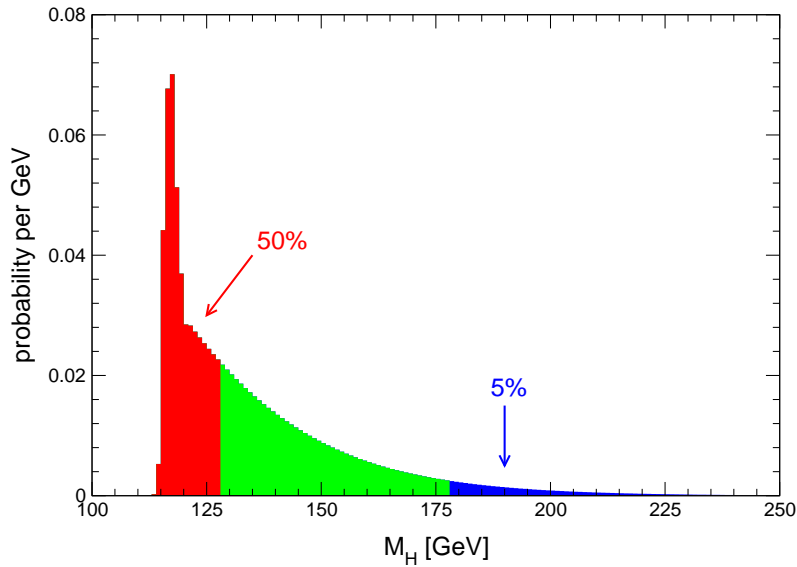}
  \caption{Probability distribution of $M_H$ from the combination of direct 
           (search) and indirect (precision) data (updated from 
           Ref.~\cite{Erler:2000cr}).}
  \label{fig:mh}
\end{figure}

Fig.~\ref{fig:mhmt} is a summary of current electroweak precision physics. It 
shows the constraints from different types of observables on the Higgs boson 
and top quark masses, $M_H$ and $m_t$. The most important input are 
the $Z$-pole asymmetries from LEP and SLC~\cite{LEPSLD:2005em,LEPEWWG:2006}, 
shown as the dotted (brown) lines. The long-dashed (blue) lines correspond to 
the $W$-boson mass, $M_W = 80.394\pm 0.029$~GeV, from LEP~2~\cite{LEPEWWG:2006}
($e^+e^-$), UA2~\cite{Alitti:1991dk} and 
the Tevatron~\cite{Affolder:2000bp,Abazov:2002bu} ($p\bar{p}$). It is 
interesting that other (non-asymmetry) $Z$-pole measurements by themselves 
result in a finite region in the $M_H$--$m_t$ plane, shown as the closed 
(green) contour. These three types of constraints overlap in a common region 
and at $m_t$ values consistent with the Tevatron average of kinematic mass 
measurements~\cite{Brubaker:2006xn}, $m_t = 171.4 \pm 2.1$~GeV. The only 
conflicting data set is from low energies (dashed contour), driven by the NuTeV
result on deep inelastic neutrino scattering off approximately isoscalar 
nuclear targets~\cite{Zeller:2001hh} which shows a 2.7~$\sigma$ deviation in 
the effective four-Fermi coupling for neutrino interactions with left-handed 
quarks. The combination of all precision data yields the filled (red) ellipse. 
There is mounting evidence for a relatively light Higgs boson with much of 
the 90\%~C.L.\ ellipse already excluded by direct searches at 
LEP~2~\cite{Barate:2003sz}. Combining these search results with the precision 
constraints yields the histogram in Fig.~\ref{fig:mh}. The strong peak is due 
to the significant excess of Higgs-like events observed by the ALEPH 
Collaboration~\cite{Heister:2001kr}. Most of the probability is for $M_H$ 
values below 130~GeV, in perfect agreement with expectations from 
supersymmetric extensions of the Standard Model (SM). The 95\% C.L.\ upper 
limit, $M_H \leq 178$~GeV, can also be read off from Fig.~\ref{fig:mh}. In 
summary, the global fit to all data yields,
\begin{equation}
\begin{array}{rcl}
M_H &=& 84^{+32}_{-25}~{\rm GeV},\\
m_t &=& 171.4 \pm 2.1~{\rm GeV}, \\
\alpha_s(M_Z) &=& 0.1216 \pm 0.0017,
\end{array}
\label{fit}
\end{equation}
where the result for $M_H$ is only barely consistent (within 1~$\sigma$) with 
the lower limit from LEP~2~\cite{Barate:2003sz}, $M_H > 114.4$~GeV. Some 
observables show interesting deviations (at the 2--3~$\sigma$ level) from 
the SM, but the overall goodness of the global fit is reasonable, with 
a $\chi^2$ of 47.3 for 42 degrees of freedom and a probability for a larger 
$\chi^2$ of 27\%. 

When discussing future improvements for the key observables, $m_t$, 
$\sin^2\theta_W^{\rm eff.}$, and $M_W$, it is useful to keep some benchmark 
values in mind. An increase of $M_H$ from 100 to 150~GeV (distinguishing 
between these values provides a rough discriminator between minimal 
supersymmetry and the SM) is equivalent to a change in $M_W$ by 
$\Delta M_W = - 25$~MeV. But this 25~MeV decrease can be mimicked by 
$\Delta m_t = - 4$~GeV, and also by an increase of the fine structure constant 
at the $Z$ scale, $\Delta\alpha(M_Z) = + 0.0014$. We know $\alpha(M_Z)$ 
an order of magnitude better than this --- despite hadronic uncertainties in 
its relation to the fine structure constant in the Thomson limit. On the other 
hand, improving $m_t$ will be important. The same shift in $M_H$ is also 
equivalent to $\Delta\sin^2\theta_W^{\rm eff.} = + 0.00021$, which in turn can 
be mimicked by $\Delta m_t = - 6.6$~GeV or by $\Delta\alpha(M_Z) = + 0.0006$. 
Thus, $\sin^2\theta_W^{\rm eff.}$ is more (less) sensitive to $\alpha(M_Z)$ 
($m_t$) compared to $M_W$, demonstrating complementarity and underlining 
the general advantage of having a diverse portfolio of measurements at ones 
disposal. Once the Higgs boson has been discovered and its mass determined 
kinematically, these observables are then free to constrain heavy new particles
which cannot be produced or detected directly. An example is the mass of
the heavier top squark in the minimal supersymmetric SM~\cite{Erler:2000jg}.

\section{LHC}
The Large Hadron Collider (LHC) is well on its way to produce first collisions 
in 2007~\cite{Rodrigo:2006}. Initial physics runs are scheduled for 2008 with 
several ${\rm fb}^{-1}$ of data and the precision program can be expected 
to take off in 2009. The low luminosity phase with about 10~fb$^{-1}$ of data 
(corresponding to 150 million $W$ bosons, 15 million $Z$ bosons, and 11 million
top quarks) per year and experiment~\cite{Pralavorio:2005ik} will already allow
most precision studies to be performed. Some specific measurements, most 
notably competitive results on $\sin^2\theta_W^{\rm eff.}$, will probably have 
to wait for the high luminosity phase with ${\cal O}(100~{\rm fb}^{-1})$ per 
year and experiment. The determination of the Higgs self-coupling would even 
call for a luminosity upgrade by another order of magnitude. Good knowledge of 
the lepton and jet energy scales will be crucial. Initially these will be known
to 1\% and 10\%, respectively, but with sufficient data one can use the $Z$ 
boson mass for calibration, allowing 0.02\% and 1\% determinations. 
Furthermore, a 2\% measurement of the luminosity and 60\% $b$-tagging 
efficiency can be assumed~\cite{Pralavorio:2005ik}.

LEP and SLC~\cite{LEPSLD:2005em} dominate the current average
$\sin^2\theta_W^{\rm eff.}= 0.23152\pm 0.00016$. Via leptonic forward-backward 
(FB) asymmetries, the Tevatron Run~II is expected to add another combined 
$\pm 0.0003$ determination~\cite{Baur:2005rx}, competitive with the most 
precise measurements from LEP (the FB asymmetry for $b\bar{b}$ final states) 
and SLD (the initial state polarization asymmetry). Having $p\bar{p}$ 
collisions are a crucial advantage here. At the LHC, by contrast, one has 
to focus on events with a kinematics suggesting that a valence quark was 
involved in the collision and which proton provided it ($Z$ rapidity tag). 
This will be possible for a small fraction of events only, requiring high 
luminosity running. Furthermore, sufficient rapidity coverage of $|\eta| < 2.5$
will be necessary for even a modest $\pm 0.00066$ 
determination~\cite{Quayle:2004ep}. A breakthrough measurement at the LHC with 
a statistical error as small as $\pm 0.00014$~\cite{Baur:2005rx,Quayle:2004ep} 
is ambitious and will require a much more challenging rapidity coverage of 
$|\eta| < 4.9$ for jets and missing transverse energy. Thus, it is presently 
unclear what the impact of the LHC on $\sin^2\theta_W^{\rm eff.}$ will be. 

At hadron colliders $M_W$ is determined by kinematic reconstruction. All 
channels and experiments combined, the Tevatron Run~II will likely add 
a $\pm 30$~MeV constraint to the world average. The huge number of $W$ bosons 
will enable the LHC to provide further $\pm 30$~MeV measurements per experiment
and lepton channel ($e$ and $\mu$) for a combined $\pm 15$~MeV uncertainty (it 
is assumed here that the additional precision that can be gained by cut 
optimization is compensated approximately by common systematics). This kind of 
measurement is limited by the lepton energy and momentum scales, but these can 
be controlled using leptonic $Z$ decays. With the even larger data samples of 
the high luminosity phase, one may alternatively consider the $W/Z$ transverse 
mass ratio, opening the avenue to a largely independent measurement with 
an error as low as $\pm 10$~MeV~\cite{Baur:2005rx}, for a combined uncertainty 
about three times smaller than our benchmark of $\pm 25$~MeV. 

The sensitivity of the total $W$ decay width, $\Gamma_W$, to new physics and 
its complementarity to and correlation with other quantities depends on how it 
is obtained. It can be extracted indirectly through measurements of cross 
section ratios,
$$
  \Gamma_W (\mbox{indirect}) = \left[ 
  {\sigma(pp \to Z \to \ell^+ \ell^- X) \over \sigma(pp \to W \to \ell \nu X)}
  \right]_{\rm exp.} \times
  \left[ 
  {\sigma(pp \to W) \over \sigma(pp \to Z)}
  \right]_{\rm th.} \times
  {\Gamma_{\rm SM}(W\to \ell \nu)\over B_{\rm LEP}(Z\to \ell^+ \ell^-)},
$$
(CDF currently quotes $\Gamma_W = 2.079 \pm 0.041$~GeV~\cite{Acosta:2004uq}) 
but the leptonic $W$ decay width, $\Gamma_{\rm SM}(W\to \ell \nu)$, has to be 
input from the SM. More interesting is therefore the direct method using 
the tail of the transverse mass distribution. An average of final Tevatron 
Run~I~ and preliminary D\O~II~\cite{Tevatron:2005ij} and 
LEP~2~\cite{LEPEWWG:2006} results gives, $\Gamma_W = 2.103 \pm 0.062$~GeV. 
The final Tevatron Run~II is expected to contribute $\pm 50$~MeV measurements 
for each channel and experiment. Detailed studies for the LHC are not yet 
available, but historically the absolute error in $\Gamma_W$ at hadron 
colliders has traced roughly the one in $M_W$. If this trend carries over to 
the LHC, a $\pm 0.5\%$ error in $\Gamma_W$ may be in store. 

Some of the Tevatron Run~II results are already included in the current 
$\pm 2.1$~GeV~\cite{Brubaker:2006xn} uncertainty in $m_t$, and with 
the expected total of about 8~fb$^{-1}$ the error may decrease by another 
factor of two. The LHC is anticipated to contribute a $\pm 1$~GeV determination
from the lepton + jets channels alone~\cite{Womersley:2006pr}. The cleaner but 
lower statistics dilepton channels may provide another $\pm 1.7$~GeV 
determination, compared with $\pm 3$~GeV from the systematics limited all 
hadronic channel~\cite{Womersley:2006pr}. The combination of these channels 
(all dominated by the $b$ jet energy scale) would yield an error close to 
the additional irreducible theoretical uncertainty of $\pm 0.6$~GeV from 
the conversion from the pole mass (which is approximately what is being 
measured~\cite{Smith:1996xz}) to a short-distance mass (such as 
$\overline{\rm MS}$) which actually enters the loops. Folding this in, 
the grand total may give an error of about $\pm 1$~GeV, so that the parametric 
uncertainty from $m_t$ in the SM prediction for $M_W$ would be somewhat smaller
than the anticipated experimental error in $M_W$.

With 30~fb$^{-1}$ the LHC will also be able to determine the CKM parameter, 
$V_{tb}$, in single top quark production to $\pm 5\%$~\cite{Mazumdar:2005th} 
(one expects $\pm 9\%$ from the Tevatron\footnote{After the conclusion of 
this conference, there was an announcement by the D\O\ 
Collaboration~\cite{Abazov:2006gd} of the first evidence of single top quark 
production. This translates into the bound $|V_{tb}| > 0.68$ (95\% C.L.).}). 
Anomalous flavor changing neutral current decays, $t\to Vq$ (where $V$ is 
a gluon, $\gamma$, or $Z$, and $q\neq b$), can be searched for down to 
the $10^{-4}-10^{-5}$ level~\cite{Womersley:2006pr}. This sensitivity gain by 
three orders of magnitude over current HERA bounds~\cite{Ferrando:2004am}, will
be relevant, {\em e.g.\/}, for extra $W'$ bosons. Measuring $t\bar{t}$ spin 
correlations at the 10\% level~\cite{Womersley:2006pr} will allow to establish 
the top as a spin 1/2~particle, to study non-standard production mechanisms 
({\em e.g.\/} through resonances), and to discriminate between $W^+ b$ and 
charged Higgs ($H^+ b$) decays. 

\begin{table}
\begin{tabular}{lccrr}
\hline
  & \tablehead{1}{r}{b}{fb$^{-1}$ per experiment}
  & \tablehead{1}{r}{b}{value [GeV]}
  & \tablehead{1}{r}{b}{error/{\sl goal}}
  & \tablehead{1}{r}{b}{$\sqrt[4]{L}$-scaling} \\
\hline
Tevatron Run I     &    0.11 &    80.452 &\bf 59 &               --- \\
LEP 2              &    0.70 &    80.376 &\bf 33 &                37 \\
\bf currently      &\bf 0.81 &\bf 80.394 &\bf 29 &            \bf 36 \\
Tevatron Run IIA   &    2    &           &\sl 31 &                29 \\
Tevatron Run IIB   &    8    &           &\sl 25 &                20 \\
LHC low luminosity &   10    &           &\sl 23 &                19 \\
LHC high luminosity&  400    &           &\sl  9 &                 8 \\
ILC                &  300    &           &\sl 10 &                 8 \\
MegaW              &   70    &           &\sl 7  &4$^*$\hspace*{-4pt}\\
\hline
\end{tabular}
\caption{Results and future expectations for $M_W$. The last column 
extrapolates the Tevatron Run~I precision under the assumption that 
sensitivities scale as in background dominated types of experiments. 
The exception is MegaW which refers to a dedicated threshold scan at the ILC 
with ${\cal O}(10^6)$ $W$ pairs and is based on a $\sqrt{L}$-scaling from 
a similar scan at LEP~2. As can be seen, such scalings provide simple estimates
of future precision goals.}
\label{tab:mw}
\end{table}

If the Higgs boson exists, its production at the LHC will proceed primarily 
through gluon fusion, $gg \to H$, and/or vector boson fusion, $qq' \to H qq'$. 
Higgs couplings can generally be determined to $10-30\%$~\cite{Baur:2005rx}. 
The top Yukawa coupling is best studied in associated production, 
$pp \to t\bar{t} H$, to $20-30\%$ precision~\cite{Womersley:2006pr}. Most 
difficult proves the Higgs self-coupling, $\lambda$, whose measurement would 
need a luminosity upgrade~\cite{Denegri:2006}. With 3~ab$^{-1}$, $\lambda$ can 
be measured to $\pm 20\%$, for 150~GeV~$< M_H < 200$~GeV, while only $\pm 70\%$
precision would be possible for a lighter (and weaker coupled) Higgs 
boson~\cite{Baur:2005rx}. 

\section{ILC}
While the hadron colliders are primarily discovery machines with remarkable 
capabilities for precision studies, the $e^+ e^-$ International Linear Collider
(ILC) --- if built --- would be a precision machine par excellence. In its 
first phase of operation the ILC would operate at center of mass energies from 
about 200~GeV (the reach of LEP~2) to 500~GeV, which would allow to scan 
the top and $ZH$ threshold regions. An integrated luminosity of 500~fb$^{-1}$ 
is expected in the first 4 years of running. The baseline design includes an at
least 80\% polarized electron beam. The second phase foresees an energy upgrade
to around 1~TeV and the collection of 1~ab$^{-1}$ of data in 3--4 years. 
A relative determination of the jet energy scale is expected to within 
$\pm 0.3/\sqrt{E}$, where $E$ is the center of mass energy in GeV. Heavy quark 
tagging will also be important and is anticipated with an efficiency at 
the 50--60\% (30--40\%) level for $b$ ($c$) quarks. 

\begin{table}
\begin{tabular}{lclll}
\hline
  & \tablehead{1}{r}{b}{fb$^{-1}$ per experiment}
  & \tablehead{1}{r}{b}{experimental value}
  & \tablehead{1}{r}{b}{error/{\sl goal}}
  & \tablehead{1}{r}{b}{$\sqrt{L}$-scaling}   \\
\hline
Tevatron Run I      &   0.072  &     0.2238  & \bf 0.0050   &   ---    \\
SLC                 &   0.05   &     0.23098 & \bf 0.00026  &   ---    \\
LEP~1               &   0.20   &     0.23187 & \bf 0.00021  &   ---    \\
\bf currently       &          & \bf 0.23152 & \bf 0.00016  &   ---    \\
Tevatron Run IIA    &   2      &             & \sl 0.0008   & 0.0009   \\
Tevatron Run IIB    &   8      &             & \sl 0.0003   & 0.0005   \\
JLab                & $\vec{e}e,~\vec{e}p$ & & \sl 0.0003   & 0.00024  \\
LHC high luminosity & 400      &             & \sl 0.00014  & 0.00008  \\
ILC                 & M\o ller &             & \sl 0.00007  & 0.00007  \\
GigaZ               & 140      &             & \sl 0.000013 & 0.000016 \\
\hline
\end{tabular}
\caption{Results and future expectations for $\sin^2\theta_W^{\rm eff.}$. Based
on $\sqrt{L}$-scaling as appropriate for statistics dominated measurements, 
the last column extrapolates the Tevatron Run~I precision to future hadron 
colliders. GigaZ refers to two years of data taking with ${\cal O}(10^9)$ $Z$ 
bosons and is scaled from the LEP~1 precision. JLab refers to 
Qweak~\cite{Armstrong:2003gp} (see next Section) and the 12~GeV M\o ller 
experiment~\cite{Mack:2006}. $\sqrt{P}$-scaling from 
E-158~\cite{Anthony:2005pm} (with $P$ the beam power) is used for M\o ller 
scattering at JLab and ILC.}
\label{tab:s2w}
\end{table}

The ILC comes with a variety of add-on options. For example, one may want 
to run it in other collision modes such as $\gamma\gamma$, $e^- \gamma$, or 
$e^- e^-$. Most relevant for precision physics would be the GigaZ 
mode~\cite{Erler:2000jg}, which would allow to repeat the LEP~1 program with 
20~million $Z$ bosons daily. The physics motivation for this option is very 
high. In particular, the world's best measurements of 
$\sin^2\theta_W^{\rm eff.}$~\cite{LEPSLD:2005em} have been provided by SLD 
($\pm 0.00029$) from the left-right cross section asymmetry, $A_{LR} = A_e$, 
and the LEP groups ($\pm 0.00028$) from the forward-backward asymmetry for 
$b$-quark final states, $A_{FB} = 3/4\, A_e\, A_b$, where 
$\sin^2\theta_W^{\rm eff.}$ is extracted from $A_e$, and where $A_b$ is 
well-known within the SM. These two measurements contribute greatly to our 
current knowledge of $M_H$. However, they disagree from each other by 
3.1~$\sigma$ and it is important to resolve this discrepancy. It is conceivably
due to new physics effects in $A_b$. Assuming this, one can turn $A_{FB}$ into 
a measurement of $A_b$ (taking $A_e$ from other asymmetries) and combine this 
with a more direct measurement of $A_b$ from SLD~\cite{LEPSLD:2005em}. 
The result, $A_b = 0.899 \pm 0.013$, deviates by 2.8~$\sigma$ from the SM 
prediction ($A_b = 0.935$). GigaZ with its ultra-high rates combined with 
polarized electrons (unlike LEP~1) would be able to determine $A_b$ to 
$\pm 0.001$! Similarly, a high precision $WW$ threshold scan with 
${\cal O}(10^6)$ $W$-pairs (``MegaW'') would allow to study the $W$ boson with
unprecedented precision. See Table~\ref{tab:mw} for a summary of $M_W$ 
measurements. The option of $e^+$ polarization 
($\buildrel > \over {_\sim} 50\%$) would allow additional cross-checks (and 
thus reduction in systematic uncertainties) and to measure new kinds of 
asymmetries. For more details see, Refs.~\cite{Yamamoto:2006,Elsen:2006}. 

Even without the GigaZ option, the ILC might provide an ultra-high precision 
measurement of the weak mixing angle (more precise than the current or 
foreseeable world average) in polarized fixed target M\o ller scattering. 
This may by then be a third generation experiment, building on the E-158 
pioneering measurement at SLAC~\cite{Anthony:2005pm} (which used the 50~GeV SLC
electron beam and achieved a precision of $\pm 0.0014$ in 
$\sin^2\theta_W^{\rm eff.}$), and a potential effort (e2ePV~\cite{Mack:2006}) 
at JLab (after the 12~GeV upgrade of CEBAF) with an anticipated error reduction
by a factor of five. The ILC would then improve on JLab by an additional factor
of four. Table~\ref{tab:s2w} summarizes future prospects for 
$\sin^2\theta_W^{\rm eff.}$.

\begin{table}
\begin{tabular}{lccrr}
\hline
  & \tablehead{1}{r}{b}{fb$^{-1}$ per experiment}
  & \tablehead{1}{r}{b}{value [GeV]}
  & \tablehead{1}{r}{b}{error/{\sl goal}}
  & \tablehead{1}{r}{b}{$\sqrt[4]{L}$-scaling} \\
\hline
Tevatron Run I      &     0.11 &     178.0 & \bf 4.3 &     --- \\
summer 2005         &     0.43 &     172.7 & \bf 2.9 &     3.1 \\
\bf currently       & \bf 1    & \bf 171.4 & \bf 2.1 & \bf 2.5 \\
Tevatron Run IIA    &     2    &           & \sl 2.0 &     2.1 \\
Tevatron Run IIB    &     8    &           & \sl 1.2 &     1.5 \\
LHC low luminosity  &    10    &           & \sl 0.9 &     1.4 \\
LHC high luminosity &   400    &           & \sl 0.7 &     0.6 \\
ILC                 &   300    &           & \sl 0.1 &     --- \\
\hline
\end{tabular}
\caption{Results and future expectations for $m_t$. The last column 
extrapolates the Tevatron Run~I precision assuming that sensitivities scale as 
in background dominated types of experiments. At hadron colliders, 
a $\pm 0.6$~GeV theory uncertainty has to be added.}
\label{tab:mt}
\end{table}

Spectacular improvements (see Table~\ref{tab:mt}) would be possible for $m_t$,
because the aforementioned short-distance mass can be extracted directly from 
the top threshold scan. 

Higgs production at the ILC would dominantly proceed through Higgs-strahlung,
$e^+ e^- \rightarrow ZH \rightarrow e^+ e^- H \hspace{4pt} (\mu^+ \mu^- H)$.
Associated production, $e^+ e^- \rightarrow t\bar{t}H$, will be important for
measurements of the top Yukawa coupling. It will be possible to determine 
the Higgs couplings $Hb\bar{b}$, $Ht\bar{t}$, $H\tau^+\tau^-$, $HW^+W^-$, and
$HZZ$ to high precision, while the $Hc\bar{c}$ will be less precise. 
The self-coupling $\lambda$ can be obtained to $\pm 20\%$ (for $M_H = 120$~GeV)
by using, {\em e.g.\/}, the process $e^+ e^- \rightarrow ZHH$. The total Higgs
width would be known to 5\%.

\section{Perspective: low energy precision measurements}
Fixed-target M\o ller scattering was already mentioned above, both at high and
relatively low energy, but all at very low momentum transfer,
$\sqrt{Q^2} = {\cal O}(100 \mbox{ MeV})$. A similar low-$Q^2$ experiment in 
elastic $e^- p$ scattering at JLab will determine the so-called weak charge of
the proton, $Q_W^p$. With an expected polarization of $85\pm 1\%$ the Qweak 
Collaboration~\cite{Armstrong:2003gp} anticipates to measure the parity 
violating asymmetry, $A_{PV} \propto (Q^2 Q_W^p + Q^4 B)$. The $Q^4 B$ term is 
the leading form factor contribution and will be determined experimentally. 
The anticipated errors in $Q_W^p$ and the corresponding $\sin^2\theta_W$ are 
$\pm 0.003$ and $\pm 0.0007$, respectively (for the calculation of the SM 
prediction, see Refs.~\cite{Marciano:1982mm,Erler:2003yk}). 

One can also explore the kinematic regimes of quasi-elastic (QE) and deep 
inelastic scattering (DIS). The discrepancy in $\nu$-DIS 
(NuTeV~\cite{Zeller:2001hh}) has already been mentioned. The interpretation of 
the experiment is hampered, however, by a variety of theoretical issues. 
An $eD$-DIS experiment is approved at JLab~\cite{Zheng:2006} to use the current
6~GeV CEBAF beam and to repeat the historical SLAC 
experiment~\cite{Prescott:1979dh} with greater precision. One hopes to be able 
to collect additional data points after the 12~GeV CEBAF 
upgrade~\cite{Reimer:2006}. This would improve the SLAC result and the current 
world average on the combination of effective four-Fermi quark-lepton
couplings $2\, C_{2u} - C_{2d}$ by factors of 54 and 17, respectively. 
The issues to be addressed are higher twist effects and charge symmetry 
violating (CSV) parton distribution functions. Since higher twist effects are 
strongly $Q^2$ dependent and CSV should vary with the kinematic variable, $x$, 
while contributions from beyond the SM would be kinematics independent, one can
separate all these possible effects by measuring a large array of data points. 
Thus, a great deal can be learned about the strong and weak interactions at 
the same time. 

Determinations of $\sin^2\theta_W$ from $\nu e$-scattering are not competitive 
at present ($\pm 0.008$), but may be another future direction, in particular if
$\beta$-beams or a $\nu_\mu$-factory became available with well-known 
$\nu$-flavor compositions and energy spectra. For example, a $\nu_e(\bar\nu_e)$
$\beta$-beam could provide a determination of $\sin^2\theta_W$ to $\pm 0.0008$ 
($\pm 0.0005$)~\cite{deGouvea:2006cb}, while a $\nu_\mu (\bar\nu_\mu)$ factory 
could achieve a precision of $\pm 0.0001$ 
($\pm 0.0003$)~\cite{deGouvea:2006cb}. Other electroweak measurements may also 
be possible at these facilities with high precision.

It should be stressed that these and other low energy tests of the SM 
(including leptonic anomalous magnetic moments, atomic parity violation, and
many SM forbidden or highly suppressed processes) will remain very important
even in the LHC era and beyond, because they are not only competitive with but
also complementary to high energy experiments. The complementarity refers to
experimental uncertainties, as well as to theoretical issues, and the way new
physics may enter. Low energy experiments may also play a prominent role in
deciphering what may be discovered at the energy frontier, even if no 
significant SM deviations are seen at low energies. 


\begin{theacknowledgments}
It is a pleasure to thank the organizers for the invitation to this meeting. 
I am particularly grateful to Miguel \'Angel P\'erez for his support and help. 
This work was supported by contracts CONACyT (M\'exico) 42026--F and 
DGAPA--UNAM PAPIIT IN112902.
\end{theacknowledgments}



\bibliographystyle{aipproc}   

\bibliography{sample}

\IfFileExists{\jobname.bbl}{}
 {\typeout{}
  \typeout{******************************************}
  \typeout{** Please run "bibtex \jobname" to optain}
  \typeout{** the bibliography and then re-run LaTeX}
  \typeout{** twice to fix the references!}
  \typeout{******************************************}
  \typeout{}
 }


\end{document}